\documentclass[aps,prl,twocolumn,superscriptaddress]{revtex4}

\usepackage{graphicx}
\usepackage{subfigure}

\def\be{\begin{equation}}
\def\ee{\end{equation}}
\def\bea{\begin{eqnarray}}
\def\eea{\end{eqnarray}}

\newcommand \Ua {\Uparrow}
\newcommand \Da {\Downarrow}

\newcommand \lan {\langle}
\newcommand \ran {\rangle}

\begin{document}

\title{Anomalous percolation and quantum criticality in diluted rare-earth
nickelates}

\author{J.V. Alvarez}
\affiliation{Department of Physics, University of Michigan.
         Ann Arbor 48109, MI, USA.}
\author{H. Rieger}
\affiliation{Theoretische Physik, Universit\"at des Saarlandes,
         66041 Saarbr\"ucken, Germany}
\author{A. Zheludev}
\affiliation{Condensed Matter Sciences Division, Oak Ridge
National Laboratory, Oak Ridge, TN 37831-6393, USA.}

\date{\today}

\begin{abstract}
A microscopic model for the diluted spin-mixed compounds
(R$_{x}$Y$_{1-x}$)$_{2}$BaNiO$_{5}$ (R=magnetic rare-earth) is
studied using Quantum Monte Carlo (QMC). The ordering
temperature is shown to be a universal function of the impurity
concentration $x$ and the intrinsic Ni-chain correlation length. An
effective model for the critical modes is derived. The possibility
of a quantum critical point driven by the rare earth-concentration
and the existence of a Griffiths phase in the high dilution limit
is investigated. Several possible experimental approaches to
verify the results are put forward.
\end{abstract}


\maketitle


An increasing number of low-dimensional quantum magnets are being
realized in materials artificially modified at nanometer scales.
The use of layered structures, dilution and/or combination of
magnetic species with different magnetic moment establish complex
geometries that enhance the role of quantum correlations. The
resulting competing ground states, quantum critical points, and
coexistence of classical and quantum features are of great current
interest in condensed matter and inter-disciplinary physics. The
family of  mixed-spin quantum antiferromagnets
(R$_{x}$Y$_{1-x}$)$_{2}$BaNiO$_{5}$ (R = magnetic rare earth)
provides a unique experimental basis to discuss these issues. The
$x=0$ compound is an excellent realization of the 1-dimensional
(1D) $S=1$ Heisenberg model, thanks to a linear-chain arrangement
of the magnetic Ni$^{2+}$ ions. Neutron scattering experiments
\cite{YBANIO} confirmed a spin liquid behavior, and revealed a
Haldane gap of about 10~meV in the magnetic excitation spectrum.
Inter-chain interactions, negligible for $x=0$, can be controlled
by substituting the intercalated non-magnetic Y$^{3+}$ by magnetic
rare earths. For $x=1$  the system orders antiferromagnetically at
$T_{N}$=10-50 K, depending on R \cite{RBANIO}. The spin dynamics
is characterized by a coexistence and frequency-separation of
semi-classical (spin waves) and quantum (Haldane-gap modes)
excitations  \cite{Z1,Z2,Z3}. By now, this phenomenon has been
extensively studied experimentally \cite{Z1}, analytically
\cite{Z2,Z3}, and numerically \cite{Z4,ALVAREZ02}.

The focus of the present Letter, however, is the role of
randomness and disorder for $0<x<1$. Previous studies were aimed
at understanding the large-$x$ limit. Due to a finite intrinsic
correlation length in Haldane spin chains ($\xi_{ch} \approx 7$
lattice units), for $x\approx 1$ disorder is effectively averaged
out. The problem is then merely a rescaled version of $x=1$ case
\cite{Z1,Z2}. For $x\lesssim 1/\xi_{ch}$ the mean distance between
magnetic rare earth centers becomes comparable to $\xi$. To date,
very little is known of this regime, beyond the fact that it
should represent qualitatively new physics. Below we develop a
numerical approach to this interesting problem. Our calculations
predict an unusual percolation mechanism for
(R$_{x}$Y$_{1-x}$)$_{2}$BaNiO$_{5}$  and related systems. We show
that the ordering temperature is a {\it universal} function of
$x\xi_{ch}$. In addition, we derive an effective model for the
critical modes, discussing whether the system is ordered down to
zero concentration $x$ or can can support a quantum critical
point.

A microscopic model for the (R$_{x}$Y$_{1-x}$)$_{2}$BaNiO$_{5}$
was proposed in Refs.~\cite{Z2,ALVAREZ02}, and studied in detail
for $x=1$ \cite{ALVAREZ02}. The corresponding spin Hamiltonian
includes a part that describes the individual Haldane spin chains
and an additional inter-chain interaction term. The former
contains no disorder and is independent of the doping level $x$.
In contrast, the interaction term involves spin operators for the
intercalated rare earth ions, and directly reflects the effect of
dilution. In the present work we will consider two variations of
this model:
\begin{eqnarray}
H_1 & = & H_{\rm c-c}
      + J \sum_{ij} {\bf S}_{i, 2j}{\bf S}_{i+1,2j}
\label{Hamiltonian_S_1}
\\
H_2 & = & H_{\rm c-c}
      - J \sum_{ij} S^{z}_{i,2j}S^{z}_{i+1,2j}+\Gamma\sum_{ij}S^{z}_{i,2j}
\label{Hamiltonian_ITF}\\
 & {\rm with} &
H_{\rm c-c} =
J_{c}\sum_{ij}\epsilon_{i,2j+1} s^{z}_{i,2j+1}(S^{z}_{i,2j}+S^{z}_{i,2j+2})
\nonumber
\end{eqnarray}
Both Hamiltonians have a common inter-chain interaction term
$H_{\rm c-c}$. $\epsilon$ is a random variable representing the
stochastic occupation of the inter-chain sites with moments
$s_{i,2j+1}$ corresponding to the rare earths with probability
density
$\rho(\epsilon)=x\delta(\epsilon-1)+(1-x)\delta(\epsilon)$. As
argued in Refs.~\cite{Z2,ALVAREZ02}, Kramers rare earth ions in
the (R$_{x}$Y$_{1-x}$)$_{2}$BaNiO$_{5}$ structure can be modeled
by $S=1/2$ pseudo-spins with anisotropic (Ising-type) R-Ni exchange
interactions $J_c$.

The two Hamiltonians differ in their description of the gaped
spin chains. In model (\ref{Hamiltonian_S_1}) the chains are S=1
Heisenberg systems, as in the real material. This model is thus
well suited for calculations of realistic ordering temperatures
and direct comparisons with experimental data. However, model
(\ref{Hamiltonian_ITF}) allows more flexibility for a conceptual
study of the problem. It incorporates quantum Ising spin chains in
a transverse field (ITF) with $\Gamma>J$ (paramagnetic gaped
regime) \cite{PFEUTY70}. The ITF model is in the same universality
class as $S=1$ Haldane spin chains, and shares with it many common
features \cite{GOMEZ-SANTOS89}. The ground state in both cases is a spin
liquid with exponentially decaying correlations with the same
asymptotic power-law corrections. The spectrum is gaped, and the
low-lying excitations are long-lived magnon-like state with a
dispersion minimum near the 1D AF zone-center. Though the
symmetries of ITF and Haldane spin chains are different (the full
$SU(2)$ symmetry of the Heisenberg model is reduced to $Z_{2}$ in
the ITF), it matters little in our case. Indeed, in the both {\it
complete} Hamiltonians \ref{Hamiltonian_S_1} and
\ref{Hamiltonian_ITF} $SU(2)$ symmetry is {\it explicitly broken}
by the Ising nature of $J_{c}$. The advantage of using the ITF
model for our purpose is that it allows to tune the intrinsic
in-chain correlation length $\xi_{ch}=\log{(\frac{J}{\Gamma})}$
through the entire range between  the free-spin limit $\xi_{ch}=0$
to chain-criticality $\xi_{ch} \rightarrow \infty$. This, in turn,
enables a study of the fine interplay between the doping level $x$
and the correlation length, and the resulting unique  percolation
behavior.

\begin{figure}
\includegraphics[width=6.0cm]{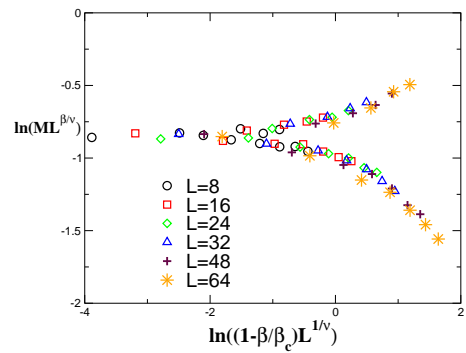}
\caption{ Finite-size analysis of the order parameter
in the vicinity of the critical point for
$J_{c}=0.1$, $\Gamma=1.25$, $x=0.5$ and $T_{c}=0.13$ in the randomly diluted
system. Here the data collapse is achieved with
$\beta=\frac{1}{8}$ and $\nu=1$, the exponents of the 2D
Ising model.
  \label{Crit_exp}}
\end{figure}

Using an efficient Quantum Monte Carlo (QMC)
cluster algorithm\cite{RIEGERQMC}\cite{EVERTZ},
%
%
\begin{figure}
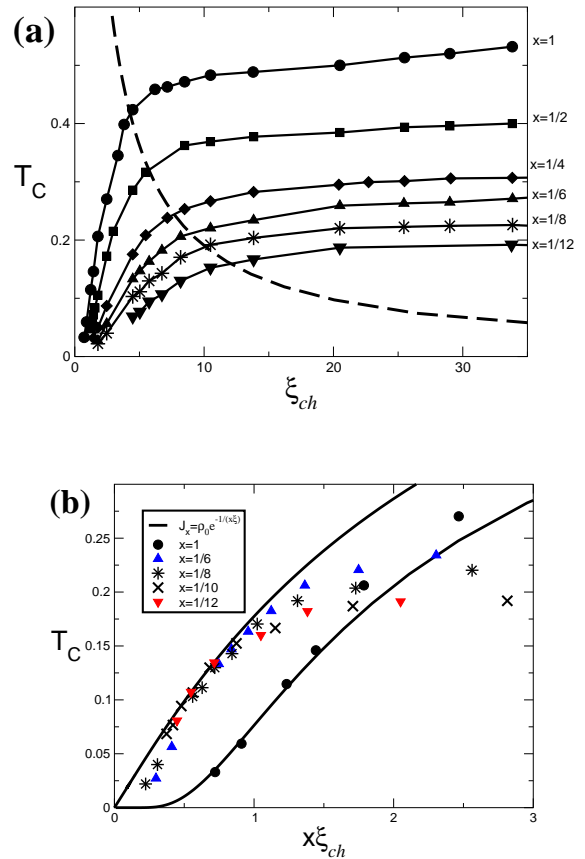

\includegraphics[width=7.5cm]{crossover.eps}

\vspace*{0.9cm}

\includegraphics[width=6.5cm]{Tc-xxi.eps}
\caption{ a.) Critical temperature $T_{c}$ as a function of the
correlation length in the {\em independent} ITF chain $\xi_{ch}$,for
different concentrations of Nd$^{3+}$ $x$.  The dashed line indicates
the value of the gap of the independent chain ($k_{B}=\hbar=1$). This 
energy scale separates the two completely different behaviors at
$T_{c} = \Delta_{ch}$ showing that the 1D features of the gaped
chains survive even at the critical point.  b.) Universal dependence
of the critical temperature on the dimensionless variable $x\xi_{ch}$.
At moderate transverse fields there is a sizable range where the
$T_{c}$ for different dilutions collapse to the critical temperature
of the 2D anisotropic Ising model (solid line).
\medskip
\label{T_c-xxi}
}
\end{figure}
we have first computed the
critical temperature for a large range
of transverse fields and concentrations in
(\ref{Hamiltonian_ITF}) using a periodic dilution pattern.
In fig. (\ref{T_c-xxi}a.)
we plot, for different concentrations,
the critical temperature $T_{c}$ as a function of $\xi_{ch}$
A sharp crossover takes places when the temperature is of the order
of the gap of the independent chain $T=\Delta_{ch}=2(\Gamma-J)$
(dashed line) considered as as a separate entity inside the whole
system. It is remarkable that a purely quantum one-dimensional
length scale is controlling the classical, two-dimensional
transition at finite temperature.
The key to understand the percolation mechanism in this system is the
universal dependence of the critical temperature
in the dimensionless variable $x\xi_{ch}$, which is the ratio between
the correlation length in the chain $\xi_{ch}$ and the average
distance between Nd$^{3+}$  ions $x^{-1}$.
For moderate values of the transverse magnetic field
different concentrations curves collapse when $x$ approaches
the low dilution limit  (see fig. \ref{T_c-xxi}b.)
This suggests that the effective degrees of freedom
that become critical interact with an effective coupling
that decreases with the distance between the Nd$^{3+}$
magnetic moments and the correlation length in the chains.
To clarify the nature of such degrees of freedom we computed
the critical exponents using the conventional finite-size analysis
of the order parameter (see fig.\ref{Crit_exp}).
We found that the exponents are compatible with
the {\em classical} 2D Ising model as well for periodic as for disordered
dilution (note that the critical behavior of the randomly diluted
2D Ising model differs from the one of the pure 2D Ising model only by
logarithmic corrections). This suggests that the building blocks
of the effective critical model are classical Ising spins.

%
%
%
%
To express this ideas in quantitative form  we have derived
a low energy  effective model for both hamiltonians
using a canonical transformation.
The central idea underlying this effective model is that
the local moments generated by the presence
of the R ion between two chains  are coupled
due to the stiffness of the finite chains connecting them.
We first study the spectrum of the local
hamiltonian of the Ni-R-Ni three-spin block:
$h_j=J_{c} s^{z}_{i,j}(S^{z}_{i, j-1}+S^{z}_{i, j+1})
+\Gamma(S^{x}_{i, j-1}+S^{x}_{i,j+1})$.
The $z$-component of the central spin $s^{z}$
representing the magnetic moment in the R$^{3+}$ ion
is a good quantum number. That symmetry breaks the block spectrum in two
orthogonal sectors. The ground state is two-fold
degenerated and each ground state belongs to distinct sectors. We label
these states  $|$$\Uparrow\rangle$ and  $|$$\Downarrow\rangle$
that have opposite values for the  expectation value of the {\em total}
$z$ component of block spin
$M^z=\pm (s^{z}+\bar{S})$ where
$\bar{S}=\mu(1+\mu^2)^{-\frac{1}{2}}$ is the
expectation value of the two outer spin magnetizations and
$\mu=J_c/2\Gamma$.
There is always a finite gap
$\Delta_{B}=((J_{c}/4)^2+\Gamma^2)^{1/2}$ to the first
excited state (level crossing of local states is forbidden by the symmetry).
The local gap does not introduce new restrictions in the validity
of our arguments because $\Delta_{B}>\Delta_{ch}$
and we are interested only in the regime  $T<\Delta_{ch}$.
All the formulas are valid also for the S=1 system taking
$J_{c}\to2J_{c}$ and $\Gamma\to0$

\begin{figure}
\includegraphics[width=7.0cm]{Tc_x.eps}
\includegraphics[width=7.0cm]{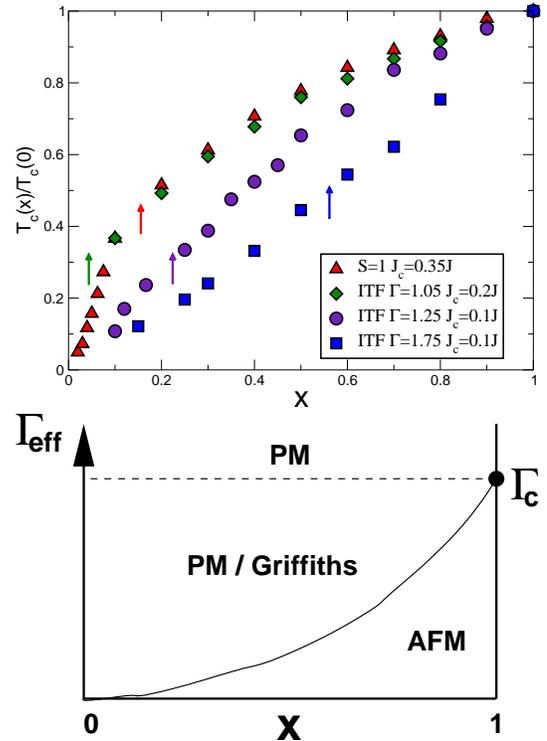}
\caption{ {\bf (a)} Normalized  critical temperature $T_c(x)/T_c(x=1)$ for
different models considered in the text.
The arrows show the value of the inverse correlation length
for each model. {\bf (b)} Generic zero-temperature phase
diagram of the randomly
diluted system described in the text. The paramagnetic region PM is
a Griffiths-McCoy region. Note that the phase boundary of the
ordered phase approaches $\Gamma_{eff}=0$ at $x=0$ with an essential
singularity. The value of $\Gamma_c$ corresponds to the QCP of the
pure system (see text).
\label{PH}}
\end{figure}

We consider now the dominant couplings in the low dilution limit:
those between two effective local moments on neighboring sites
along the direction of the chains $i,i+1$.
For model (\ref{Hamiltonian_ITF}) this coupling is exclusively
of the Ising type (there are no spin-flip process along the chains).
The leading contribution to this coupling is given by the energy
necessary to create a domain wall between blocks $i$ and $i+1$,
%
%
\[
J_{\parallel}=\frac{1}{2}(\lan \Ua_{i},\Da_{i+1}
|h^{z}_{\parallel}|\Ua_{i},\Da_{i+1} \ran -\lan \Ua_{i+1},\Ua_{i}
|h^{z}_{\parallel}|\Ua_{i},\Ua_{i+1} \ran) \]
%
%
where
$h^{z}_{\parallel}=J(S^z_{i,j-1}S^z_{i+1,j-1}+S^z_{i,2}S^z_{i+1,j-1})$.
Using the explicit form of the block eigenstates
we can proof that the effective coupling is
$J_{\parallel}=J \bar{S}^2 +O(J^2/\Delta_B)$,  i.e. the product
of the energy to create a domain wall in each chain
times the expectation value of the magnetization in the two
outer spins of the block.
In the case of $S=1$ Heisenberg chains
we have off-diagonal terms
$h^{xy}_{\parallel}=J/2(S^{+}_{i,j-1}S^{-}_{i+1,j-1}+S^{+}_{i,j+1}S^{-}_{i+1
,j+1})+$ h.c..
However, the $s^z_{i}$  are still good quantum numbers
and this prevents spin-fliping in the effective Hamiltonian.
This is the strongest argument in favor of the low-energy equivalence
of models (\ref{Hamiltonian_S_1}) and (\ref{Hamiltonian_ITF}).
An essentially identical argument can be applied in the high
dilution limit  to compute effective coupling between two distant
local magnetic moments obtaining now $J_{\parallel}=2\bar{S}^2
\rho$ where
$\rho=E(s_j=s,s_k=-s)-E(s_j=s,s_k=s)$
$E(s_j,s_k)$ is the energy of the chain
connecting the two blocks with spin $s_j$
in site $j$  and  $s_k$ in site $k$. In other words, $\rho$ is the energy
difference between anti-parallel and parallel spins
at the end of the chain and it is directly related
to the static stiffness of the (finite) chain
which is non-zero even in the paramagnetic case.
In both chain models the asymptotic behavior is
$\rho \sim  \rho_{0}\exp(-|k-j|/\xi_{ch})$.
We can test numerically the validity of this model in the
periodic diluted system. The critical temperature for the 2D
anisotropic Ising model is given by the expression $\sinh{\beta_c
J_{x}} \sinh{\beta_c J_{y}}=1$ where $J_{x}=J_{c}\bar{S}$ and
$J_{y}=2J^{2}_{c}\frac{1}{\sqrt{x}}e^{-\frac{x}{\xi}}$ and
$\beta_c=1/T_{c}$ . The results (solid line) are compared with the
numerical simulation in Fig. \ref{T_c-xxi}b.
The agreement emerges at $T_{c}<<\Delta_{B}$ where
thermal population of excited states above the doublet is strongly
suppressed and the block is described by an Ising variable.  The
effect of the quantum fluctuations in the microscopic models
is integrated out in the effective classical model and
can be formally separated into two parts: the reduction of the
effective magnetic moment of the Ising spins $|$$\Ua\ran$,$|$$\Da\ran$ and
the exponential decay of the couplings in the direction parallel to
the chains.  The most prominent feature of the model in the low
dilution limit is the linear behavior of $T_{c}$ as a function of x
at fixed $\xi_{ch}$. This property survives in the more realistic
randomly diluted case for a large range of concentrations (see
Fig. \ref{PH}a).

It is interesting to consider whether
the absence of percolation threshold is a robust feature against the
presence of weak spin-flip processes between the central spin R and
its neighbors in the perpendicular direction that belong to the
chains, $h^{XY}_{\perp}= J^{xy}_{c}/2 s^{+}_{i,j}(S^{-}_{i,
j-1}+S^{-}_{i,j+1})+$ h.c. . A finite value for $J^{xy}_{c}$ breaks the
conservation law for $s^{z}_{i}$ and establishes quantum coherence between
$|$$\Uparrow\rangle$ and $|$$\Downarrow\rangle$. In leading order, it has
the same effect of a transverse magnetic field acting on the spin blocks.
If $J_{XY}<<\Delta_{L}$ we can compute within second order perturbation
theory the matrix element $\Gamma_{\text{eff}}=\langle
\Downarrow \left| H_{xy} \right| \Uparrow \rangle=J_{XY}^{2}/4\Delta_{L}$.
This $\Gamma_{eff}$
should not be confused with the original transversal field $\Gamma$ in
the chains. The effective model at T=0 is now a 2D ITF in
that has a quantum critical point (QCP)
for $\frac{\Gamma}{J}=3.044(2)$ \cite{RIEGERQMC2}.  In the original model
this translates into a QCP governed by the
concentration and therefore a finite percolation threshold,
$x_{c}^{-1}=\xi_{ch}\log(\frac{\Gamma_{\text{eff}}}{\alpha \rho_{0}})$ where
$\alpha$ is a numerical constant. Due to the
fact that the random dilution affects only inter-chain sites 
a percolating cluster of magnetic moments exists with probability one
for any finite value of $x$. But unlike the homogeneously diluted $S=1/2$ 2D
Heisenberg systems \cite{SANDVIK}, it can be can be
disordered by pure quantum fluctuations.

In Fig. \ref{PH}b we show the generic zero temperature phase diagram
that we expect for the randomly diluted system: Since the critical
value for the effective matrix element $\Gamma_{eff}$ vanishes
exponentially for $x\to0$, strongly diluted ($x\ll1$) experimental
systems are most probably in the paramagnetic phase (PM). Nevertheless,
since the system is disordered and close to a quantum phase transition
at $\Gamma_{eff}(x)$, one expects the region $\Gamma_{eff}<\Gamma_c$
to be a Griffiths-McCoy phase \cite{griffiths,mccoy,qsgreview},
in which various
physical observables display a singular behavior and experimental
measurements should show strong sample to sample fluctuations as
we actually observed in the numerical simulations.
The usual phenomenological argument leading to, for instance, a
diverging susceptibility in a diluted transverse Ising system close to
but away from a QCP is the following
\cite{HUSE,RIEGERSG}: Connected clusters with $N$ spins appear with an
exponentially small probability $p^N=\exp(-\lambda N)$ ($p$ being the
probability for an occupied site), but first order perturbation theory
tells us that for small transverse field strengths the gap of this
cluster is also exponentially small in the number of connected spins
$\Delta\sim\exp(-\sigma N)$, leading to an exponentially large
tunneling time. This implies, at zero temperature, a power law
behavior for the autocorrelation function
$[\langle\sigma_i^z(t)\sigma_i^z(0)]_{\rm av}\sim
t^{-\lambda/\sigma}$, and thus an algebraic singularity of, for
instance, the local zero-frequency susceptibility $\chi(\omega=0)\sim
T^{-1+\lambda/\sigma}$. These Griffiths-McCoy singularities are
generic for a randomly diluted transverse Ising system (actually
generic for any kind of disorder). In our case the dilution is such
that no isolated finite clusters exist, nevertheless parts of the
system are more strongly coupled than others simply
because there are more occupied inter-chain sites. Generic disorder,
like random ferromagnetic bonds, also leads to Griffiths-McCoy
singularities in non-diluted transverse Ising systems, as was
demonstrated numerically \cite{RIEGERQMC}, hence we also expect them
to be present in the system we were considerng here.

In conclusion we would like to stress the experimental
implications of our theoretical findings. Our central predictions,
particularly the universal scaling of the ordering temperature,
should be straightforward to verify on
(R$_{x}$Y$_{1-x}$)$_{2}$BaNiO$_{5}$ powder samples by bulk
magnetic and calorimetric measurements or by neutron diffraction.
Conversely, the intrinsic chain correlation length can be
effectively {\it measured} through measuring the $x$-dependence of
the ordering temperature. The finer effects discussed above,
including those associated with strong disorder effects and algebraic
singularities, can be directly probed in more involved
single-crystal neutron scattering experiments.

J.V.A. thanks C. Gros and S. Moukouri for support. Work at ORNL
was carried out under DOE Contract No. DE-AC05-00OR22725.



\begin{thebibliography}{99}

\bibitem{YBANIO} ~J.~Darriet,~L.P.~Regnault, Solid State Commun. {\bf
86},409(1993),J.F. DiTusa et al. Physica B 194-196,181 (1994). T. Sakaguchi
et al. J. Phys. Soc Jpn {\bf 65},3025 (1996).
G.Xu et al. Phys. Rev. B {\bf 54},  R6827  (1998).

\bibitem{RBANIO} E. Garc\'{\i}a-Matres {\it et al.}, J. Solid Sate
Chem. {\bf 103}, 322 (1993); J. Magn. Mag. Mat. {\bf 149}, 363 (1995).


\bibitem{Z1} A. Zheludev {\it et al.}Phys. Rev. Lett. {\bf 80}, 3630
(1998);
Phys. Rev. B {\bf 58}, 14424 (1998); Phys. Rev. B {\bf 61}, 11601
(2000); J. Phys.: Condens. Matter {\bf 13}, R525 (2001) and
references therein.

\bibitem{Z2} S.~Maslov and ~A.~Zheludev, Phys. Rev. Lett. {\bf
80}, 5786 (1998).

\bibitem{Z3} I. Bose and E. Chattopadhyay, Phys. Rev. B {\bf 60}, 12138
(1999); E. Ercolessi {\it et al.}, Phys. Rev. B {\bf 62}, 14860
(2000).


\bibitem{Z4} J. Lou, X. Dai, S. Quin, Z. Su and L. Yu, Phys. Rev.
B {\bf 60}, 52 (1999).

\bibitem{ALVAREZ02} J. V. Alvarez, R. Valenti, A. Zheludev
      Phys. Rev. B {\bf 65}, 184417 (2002).

\bibitem{PFEUTY70} P. Pfeuty,  Ann. Phys. {\bf 57}, 70 (1970).

\bibitem{GOMEZ-SANTOS89} G. Gomez-Santos, 
  Phys. Rev. Lett. {\bf 63}, 790 (1989).

\bibitem{RIEGERQMC} C. Pich, A. P. Young, H.\ Rieger and N.\ Kawashima,
  Phys.\ Rev.\ Lett.\ {\bf 81}, 5916 (1998).

\bibitem{EVERTZ} H.G. Evertz, G. Lana and M. Marcu,
                 Phys. Rev. Lett. {\bf 70}, 875 (1993).

\bibitem{RIEGERQMC2}
  H.\ Rieger and N.\ Kawashima, Europ.\ Phys.\ J.\ B {\bf 9}, 233 (1999).

\bibitem{SANDVIK} A.W. Sandvik, Phys. Rev. B {\bf 66}, 024418 (2002).

\bibitem{griffiths} R.~B.~Griffiths, Phys. Rev. Lett. {\bf 23}. 17 (1969).

\bibitem{mccoy} B. M. McCoy, Phys. Rev. Lett. {\bf 23}, 383 (1969).

\bibitem{qsgreview} H.\ Rieger and A.\ P.\ Young,
  {\it Quantum Spin Glasses}, Lecture Notes in Physics {\bf 492}
  ``Complex Behaviour of Glassy Systems'', p.\ 254,
  ed.\ J.M. Rubi and C. Perez-Vicente. Springer Verlag (1997).

\bibitem{HUSE} M.~J.~Thill and D.~A.~Huse, Physica A, {\bf 15}, 321 (1995).

\bibitem{RIEGERSG} H.\ Rieger and A.\ P.\ Young,
  Phys.\ Rev.\ B {\bf 54}, 3328 (1996); T. Ikegami, S. Miyashita and
  H. Rieger, J. Phys. Soc. Jap. {\bf 67}, 2761 (1998).

\end{thebibliography}
\end{document}